\documentclass[12pt]{article}

\usepackage{amsmath}
\usepackage{amscd}
\usepackage{amsfonts}
\usepackage{amssymb}
\usepackage{amsthm}
\usepackage{epsfig}

\renewcommand {\d}  {\partial}
\newcommand {\vd}   {\delta}
\renewcommand {\phi}{\varphi}
\renewcommand {\a}  {\alpha}
\newcommand {\N} {\mathcal{N}}
\newcommand {\F} {\mathcal{F}}

\newcommand{\f}{\frac}

\newcommand{\res}{\mathop{\rm res}}

\newcommand{\vo}[1]{Vol(U(#1))}
\newcommand{\cint}[1]{\frac 1 {2\pi i} \oint_{#1}}

\newcommand{\ti}{\tilde}

\def\appendix#1{
  \addtocounter{section}{1}
 \setcounter{equation}{0}
  \renewcommand{\thesection}{\Alph{section}}
 \section*{Appendix \thesection\protect\indent \parbox[t]{11.715cm} {#1}}
  \addcontentsline{toc}{section}{Appendix \thesection\ \ \ #1}
  }

\newcommand{\newsection}{
\setcounter{equation}{0}
\section}

\newcommand{\eq}[1]{\begin{equation} #1 \end{equation}}
\newcommand{\ar}[1]{\begin{eqnarray} #1 \end{eqnarray}}

\newcommand{\muu}[1]{\begin{multline} #1 \end{multline}}

\newcommand{\tr}{\mathop{\mathrm{tr}}\nolimits}
\newcommand{\Tr}{\mathop{\mathrm{Tr}}\nolimits}

\def\d{\partial}
\def\D{\Delta}

\newcommand{\br}[1]{\left( #1 \right)}
\newcommand{\vev}[1]{\left\langle #1 \right\rangle}
\newcommand{\rf}[1]{(\ref{#1})}

\def\ll2{\ll}

\title{
\hfill{\small ITEP-TH-14/03}
\\~\\
On the property of Cachazo-Intriligator-Vafa prepotential at
the  extremum of the superpotential.
 }

\author{A. Dymarsky\thanks{\tt dymarsky@gate.itep.ru}~
and V. Pestun\thanks{{\tt pestun@gate.itep.ru}}
\\~~\\
{\it Institute of Theoretical and Experimental Physics} \\ {\it
B.~Cheremushkinskaya 25, 117259 Moscow, Russia}
\\~~\\
}

\date{January, 2003}
\begin{document}           

\maketitle

\abstract{
We consider CIV-DV
prepotential $\F$ for $\N=1$ $SU(n)$ SYM theory at the
extremum  of the effective
 superpotential and prove the relation
 $$2\F-S^i {\d \F \over \d S_i}=-u_2{2\Lambda^{2n}\over(n^2-1)}$$}

\newsection{Introduction}

In this note we consider the Cachazo-Intriligator-Vafa prepotential \cite{CIV,DV} for $\N=1$ $SU(n)$ SYM
theory, which is equal \cite{DV} to the free energy $F_m$ for the one matrix holomorphic
integral with the polynomial potential, defined as a perturbative
expansion around the certain extremum of the action.

\eq {
\label{def1}
e^{-{F}_m} = \lim_{N \rightarrow \infty} {\Lambda^{-N^2}\over {\vo{N}}} \int d\ti{\Phi}_{N\times N}
e^{-\frac 1 {g_s} \tr W(\Phi_{0}+\ti{\Phi})}
}
Here $\Phi=\Phi_0+\ti{\Phi}$ is a complex matrix,
and the integral is an analogue of a contour integral in the one
dimensional case, so that the real dimension of the integration domain is $N^2$.
Naively, one can treat the integral
just as an ordinary hermitian matrix integral in the case of the
even order potential, but generally speaking, one should explicitly specify
the domain of the integration for  the integral to be convergent.
While we do not study the problem in its full complexity,
we yet adopt the prescription \cite{DV} of the perturbative expansion around some extremum of the action
(it becomes an additional argument of $F_m$),
and in this case we have not to explicitly specify a contour of integration, since
the result depend only on a local structure of the action around the expansion point. So, in a perturbative
calculation one can treat the integral just as an ordinary hermitian one matrix model with $U(N)$ symmetry \cite{DV}.

The action $W(x)$ is  a polynomial of degree $n+1$. We define \eq{
W'(x) = \prod_{i=1}^{n} (x-\a_i) \equiv x^n - \sum_{k=2}^{n} g_k
x^{n-k},\qquad \sum_{i=1}^n \a_i=0 }


The critical points of the matrix action are given by specifying \cite{DV,Thei,Nac} how
$N$ eigenvalues of the matrix $\Phi_0$ are distributed around the
 critical points $\a_i$ of the potential $W(x)$.

Let $N_i$ is the number of the eigenvalues of $\Phi_0$ that are
equal to $\a_i$. In the planar (quasiclassical) limit $g_s \to
0$ the result for the free energy can be represented as an
expansion over the genus of fat Feynman diagrams for the matrix
model
\eq { \label{expan} F_{m}= \sum_{g=0}^{\infty} g_s^{2g-2}
F_g (S_i,\a_i)
}
with  $S_i=g_s N_i$.
We focus on the planar
contribution $g=0$ to the free energy $F$, from which the SW
solution for the pure $\N=2$ gauge theory  can be constructed
\cite{DV}.

It has  the following structure
\eq{ \label{F_pl}
F_{pl}=\frac {1} {g_s^2} F (S_i \equiv g_s N_i ,\a_i),\qquad F\equiv F_0 }
with the following leading terms of $F$ \cite{CIV,DV,Thei,Nac,Mor}:
\eq{
\label{F} F(S_i,\a_i) = \sum_{i} W(\a_i)
S_i - \sum_{i} \br {\frac 1 2 S_i^{2} \log {\frac {S_i}
{\Lambda^2\D_i}} - \frac 3 4 S_i^2} - \sum_{i<j} 2 S_i S_j \log
\br{\f {\alpha_{ij}} \Lambda} +O(S^3) }
where
\ar{
a_{ij} \equiv \a_i - \a_j \\
 \Delta_i \equiv \prod_{j \neq i} a_{ij} = W''(\a_i)
}
The first piece of \rf{F_pl} is the classical contribution, the second is due to the logarithm of the
volume of $U(N_i)$ group and the gaussian integration,
and the third is from the jacobian of $U(N) \to \prod U(N_i)$.

Then, the following object is believed \cite{CIV,DV,SW}  to be a superpotential in the $\N=1$ effective low energy
theory, that is obtained after integration out the adjoint field $\Phi$:
\eq{
W_{eff} (S_i,\a_i) = \sum_{i} \frac {\d F} {\d S_i}
}

According to \cite{CIV,DV}, to construct the SW solution \cite{SW_solution} for the $\N=2$ theory,
one should consider an extremum point  $\vev{S_i}$ of $W_{eff}$
\eq{\label{d2F}
\left. \frac {W_{eff}} {\d S_{i}} \right|_{\vev{S_i}} =\left. \sum_{j} \frac {\d^2 F}
{\d S_i \d S_j}\right|_{\vev{S_i}} =  0
}
since it is believed that in this point
 $f_{n-1}=-4\Lambda^{2n}$ \cite{CIV,DV} and
matrix model curve $y^2={W'}^2+f_{n-1}$ becomes the SW curve \cite{SW_solution}.

This condition  \rf{d2F} could  be easily explained from the field theory
side \cite{DV,SW}. It means decoupling of the common $U(1)$ factor in the effective low-energy theory.
Unfortunately, in the matrix integral
picture the meaning of this condition is not clear yet, in spite of the rapid development of the issue \cite{all}.
We hope that it could be interpreted as a condition for a point, where the perturbative calculation
in some "right" way (still not understood rigorously now) coincides with the exact, non perturbative
definition of the matrix integral \rf{def1} \cite{MorRev}.

In this letter we proposed(instead of discuss) another non trivial relation that holds in this point
\begin{footnote}
{but not in a general point!}
\end{footnote}
$\vev{S}$,
 involving
not only the second derivatives, but the value of $F$ and its first derivatives itself, showing
that for the matrix integral the point $\vev{S}$ is indeed a very special one!

This relation is (here $u_2 \equiv \f 1 2 \sum \a_i^2 = g_2$ due to $\sum
\a_i = 0$)
\eq{
\label{mat}
\left. V \equiv 2F - \sum S_i \frac {\d F} {\d S_i}\right|_{\vev{S_i}} = - \frac { 2 g_2\Lambda^{2n}} {n^2-1}
}
This relation was recently discovered by Matone \cite{Matone} in the $SU(2)$ case in a slightly
different form and was checked by the perturbative expansion. We
generalize and prove this relation for an arbitrary unitary gauge group.

Recall, that a similar fact about
the $\N=2$ SW effective prepotential
\begin{footnote}
{which also  was discovered by same author  \cite{M1} and
 after generalized for other groups and proved in all orders in two ways \cite{Sonnenschein,Eguchi}}
 \end{footnote}
\eq{ \label{x}
{\d F_{SW}\over \d \ln\Lambda^2}=2\,F_{SW} - \sum a_i \frac {\d F_{SW}} {\d a_i} =-8\pi i\beta_0 u_2
}has a clear RG interpretation in $\N=2$ SYM theory \cite{Bonelli:1996qc}.
Its simple form can also be explained using superconformal Ward identities
\cite{Howe}.

It will be very interesting to find a similar interpretation for the identity
\rf{mat} in terms of matrix model RG \cite{itoi,brezin} and
to explain simple and very similar to the r.h.s. of \rf{x} form of the r.h.s. of
\rf{mat}.
Probably, it is possible to find some interpretation of this relation on a
field theory side, similarly to the interpretation of the Virasoro constraints in
terms of the generalized Konishi anomaly \cite{Gor,SW}.

It should be also remarked that the relation \rf{mat} looks like a consequence
of the exactness of quasiclassical approximation.
In fact, dropping all terms except the  classical (the first) one in \rf{F}
\eq{\label{ff} F^{cl}=\sum W(\a_i) S_i^{cl}}
and using only classical
\begin{footnote}{
the leading in the expansion series over $\Lambda$}
\end{footnote} term for $\vev{S}$
\eq{\label{Scl}
S_i^{cl} = \f {\Lambda^{2n}} {\D_i}}
one could easily get the result
\eq{
V^{cl}  = \sum \f {W(\a_i)  \Lambda^{2n}} {\D_i}=\f {\Lambda^{2n}} {2\pi i} \oint \f {W(z)} {W'(z)} dz=
  - \f {2 g_2 \Lambda^{2n}} {n^2-1}
}
The relation is not the only one that holds exactly in its classical form at the special point
$\vev{S}$. The classical result for the superpotential that follows
from \rf{ff}
\eq{W_{eff}=\sum W(\alpha_i)}
is  also exact at the special point $\vev{S}$ \cite{CIV,DV}.

The property of the classical approximation to be exact is some consequence
of the planar limit in \rf{def1} and, perhaps, it can enlighten the question of the role of
the special point $\vev{S}$ from the matrix model point of view.

\section{The first proof of the  relation}
In this section we present the proof of the relation \rf{mat}.
We consider the planar contribution to the free energy that has a structure
\eq{
F_{pl} = g_s^{-2} F(g_s N_i,\a_i)
}
Differentiating over $g_s$ we see
\eq{
\d_{g_s} F_{pl} =  (-2 g_s^{-3} F + g_s^{-3} \f {\d F} {\d S_i} S_i) = -g_s^{-3} V
}
From the definition of the $F_{pl}$ as a free energy we know, that its derivative over a parameter
is a vacuum expectation value of the correspondingly coupled operator.
Since in the planar limit
\eq{
F_{pl} = -\log \br{ \int d \Phi e^{-g_s^{-1} \tr W(\Phi)}}
}
then
\eq{
-g_s^{-3} V = \d_{g_s} F_{pl} = -g_s^{-2} \vev { \tr W(\Phi)}
V = g_s \vev{ \tr W(\Phi)}
}
The expectation values can be calculated with the help of the exact algebraic equation
that can be obtained in the planar limit
 for the resolvent \cite{SW}:
\eq {
R(z) \equiv g_s \Tr \vev {\f 1 {z - \Phi}}
}

The loop equation (or Ward identity  for the variation $\vd \Phi = \f 1 {z-\Phi}$) in the planar
limit (when the correlation functions for single trace operators factorize like
$\vev{A^i_k B^k_j}=\vev{A}^i_k\vev{B}^k_j$)
 is the following
\eq{
\label{loopeq}
R(z)^2 = R(z) W'(z) + \f 1 4 f_{n-1}(z)
}
Here $f_{n-1}(z)$ is yet an arbitrary polynomial of the order $n-1$. Its $n$ coefficients
 specify values of $S_i$ through the relation
\eq{
\label{Si}
S_i = \frac 1 {2\pi i} \oint_{A_i} R(z) dz
}
Here and below we use the  notations of \cite{Cachazo:2003zk} for cycles on complex plane.
From the equation \rf{loopeq}
we read the solution $R(z) = 1/2 \br{ W'(z) - \sqrt{W'(z)^2+f_{n-1}(z)}}$.

Via $R(z)$ one can easily calculate the expectation values of the single trace operators
 $\tr \Phi^k$ like:
\ar{
g_s\vev{\tr \Phi^k} =  \frac 1 {2\pi i} \oint_{A} R(z)z^k dz\\
A=\sum_{i=1}^n A_i, \qquad   \forall ~\phi(z)    \qquad {1\over 2\pi i}\oint_{A} dz~ \phi(z)=\res_{z=\infty} \phi(z)
}

For an arbitrary polynomial we have
\eq{
g_s\vev{\tr  P(\Phi) } =  \frac 1 {2\pi i} \oint_{A} P(z) R(z) dz
}

From \cite{CIV,DV,SW} we know that at the special point $f_{n-1} =- 4\Lambda^{2n}$
and the leading term for $\vev{S}_i$ is indeed given by \rf{Scl}

\eq
{
\vev{S}_i = -\cint{A_i} \f 1 2 \sqrt{ (x+\a_i-2\Lambda^{n}/\D_i)(x+\a_i + 2\Lambda^{n}/\D_i) } \D_i +...=
{\Lambda^{2n}\over{\D_i}}
}
Now we are near the desired result
\eq{
g_s\vev{\tr W(\Phi)} = - \cint{C} W(x) {  {\sqrt {W'(x)^2-4\Lambda^{2n}} }\over 2} =
 \res_{z=\infty} \f {W(z)} {W'(z)}\Lambda^{2n}
}
because
\eq{
\res_{z=\infty} \f {W(z)} {W'(z)} = \res_{z=\infty}
 \f {\f 1 {n+1} x^{n+1}-g_2 \f 1 {n-1} x^{n-1} +..} { x^{n}-g_2 x^{n-2} +..} = g_2 \br {\f 1 {n+1} - \f 1 {n-1}} = -g_2 \f {2} {n^2-1}
}
This is the end of the first proof.

\section{The second proof of the  relation}
The second proof of \rf{mat} checks the  definition of $S_D\equiv{\partial F\over \partial
S}$ as an integral over some non closed contour $B_i$ \cite{CIV}, and determines the
r.h.s. of \rf{mat} up to an overall constant, which was
checked by the perturbative calculations in the introduction. Note, that
the coincidence between the perturbative matrix model calculation for ${\partial F\over \partial S}$
\cite{DV,Thei,Nac}
and the integral of some differential along $B_i$ cycle \cite{CIV,Mor} is not straightforward,
and a lot of work is required to check this identity explicitly even in the first few orders.

Let us consider the partial derivative of  \rf{mat} over $g_k$ considering
$S_i=\vev{S}_i$ as the functions that  depend on $g_k$ and $\Lambda$
\eq{\label{bn}
{\d V \over \d g_k}=2{\partial F\over \partial g_k}+ S^{i}_D
{\partial S^i\over \partial g_k }-S^{i} {\partial S^{i}_D\over \partial g_k }
}
at the special point  $\vev{S}$.

First term could be calculated similarly to the derivative over $g_s$:
\muu{\label{rt}
2{\partial F\over \partial g_k}=-2 \f {g_s \vev{ \Tr \Phi^{n-k+1}}} {n-k+1} ={1\over 2\pi i}
\oint\limits_{A} {dz~ z^{n-k+1}\over n-k+1} \br{y-W'}\\
=\oint\limits_{A} {dz~ \over n-k+1} \br{{ -4\Lambda^{2n}z^{n-k+1}\over
2W'}+O\br{{1\over z^{2n}}}}=-{2\Lambda^{2n} \over (n-1)}\delta_{k,2}
}
The values $S=(g_k,\Lambda)$ and $S_D(g_k,\Lambda)$  defined as
\cite{DV}
\ar{
\label{def}
S^i={1\over 2\pi i}\oint\limits_{A_i} R(x) dx\\
S_D^i=\oint\limits_{B_i} R(x) dx\\
R(x)=\f 1 2 \br{W'(x)-y(x)}
}
Note that the non-compact cycles $B_i$ has a nice property $B_i-B_n=-b_i$ for
$i=\overline{1,n-1}$, where compact cycles $b_i$
and $a_i=A_i$ for $i=\overline{1,n-1}$ form the canonical basis of cycles
on the hyperelliptic curve $y^2={W'}^2-4\Lambda^{2n}$.

On the special curve $f_{n-1}=-4\Lambda^{2n}$ the differential $ydx$ has no residues. That is why
$\sum_i \vev{S_i}=0$, and thus we can change the contour from $B_i$ to $b_i$
($b_n=0$) in the definition of $S_D$ \rf{def} without changing \rf{bn}.

Now the second and the third terms in \rf{bn} could be easily calculated with a help of the Riemann
bilinear relation \cite{Riemann} yielding
\ar{\label{er}
S^{i}_D {\partial S^i\over \partial g_k }-S^{i} {\partial S^{i}_D\over \partial g_k
}= - {1\over 8\pi i}\sum_i \br{\oint_{a_i}y dx\oint_{b_i}{\d y \over \d g_k}dx-\oint_{a_i} {\d y \over \d g_k}dx\oint_{b_i} y dx} =
\\
=-\f {1} 4 \sum~ \res \br{\int^x_{x_0} y(z) dz{\partial y\over\partial g_k}(x)}}

It is useful to set $x_0$ to coincide with some root of $y(x)$,
then this point will belong to the both branches of the curve, and
thus the eventual result is twice larger than the one pole result.

At the infinity (near the pole)
\eq{\label{q1}
\int^x_{x_0} y(z) dz=W(x)+{4\Lambda^{2n}\over 2(n-1) x^{n-1}}+\mathrm
{const}+ O\br{{1\over x^{n}}},~~x \to \infty
}
and
\eq{\label{q2}
{\partial y\over\partial g_k}(x)=-x^{n-k} \br{1+{4\Lambda^{2n} \over 2 (W')^2}+O\br{{1\over x^{4n}}}}~~x \to \infty
}
Multiplying these two quantities by each other one could simply  get
\eq{\label{we}
\int^x_{x_0} y(z) dz{\partial y\over\partial g_k}(x)=-{4\Lambda^{2n}\over 2 x^{k-1}}\br{{1\over (n+1)}+{1\over (n-1)}}+Q(x)+O({1\over
x^{k+2}}),~~ x\to \infty
} where $Q(x)$ is some polynomial.
Thus, substituting \rf{we} into \rf{er} and adding it to \rf{rt}we get
\eq{\label{r}
{\d V\over \d g_k}=-{\Lambda^{2n}}\br{{1\over (n-1)}-{1\over (n+1)}}\delta_{k,2}
}
This is the end of second proof, since the result \rf{r}  coincides with the partial
derivative over $g_k$ of \rf{mat}.

\section{Conclusion}

In this note we proposed a new relation which holds for the CIV prepotential
at the special $\N=2$ point $\vev{S}$.
As was mentioned above this relation has the similar form to the renormalization group equation
for the Seiberg-Witten prepotential,
 and it would be interesting to understand it from the matrix model RG point of view.

 We  expect the straightforward generalization of this relation for other
gauge groups \cite{DVSoSp} and for theory with matter \cite{Seiberg:2002jq}.

We hope that the proposed relation will help to specify in some natural way
the conditions for the point $\vev{S}$ at the matrix model side, and thus
gives more evidence on relation between matrix models and
SUSY gauge theories.

\section{Acknowledgments}
We would like to thank  A. Gorsky, A. Mironov and especially A.~Morozov
for helpful discussions, suggestions and comments.
We acknowledge for the kind hospitality the 20th Jerusalem Winter School on "String Theory:
From Confinement to Cosmology", January 2003, held at the Institute for Advanced Studies, the Hebrew
University of Jerusalem, where this work was partly done.

This work was partly supported  by the RFBR grant 01-02-17682 (A.D.),
INTAS grant 00-334 (A.D.),  the RFBR grant
01-02-17488 (V.P.), the INTAS grant 00-561 (V.P.), and by the Russian
President's  grant 00-15-99296.

\end{document}